\documentclass[12pt,a4paper]{article}
\usepackage[pctex32]{graphics}
\usepackage{amssymb,amsmath}
\begin{document}

\title{\large \bf
Theory of amplification of sound (acoustic phonons) by absorption of laser radiation in quantum wires with parabolic potential.}
\author{\normalsize Nguyen Quoc Hung, Dinh Quoc Vuong, Nguyen Quang Bau\\
\normalsize \it Faculty of Physics, Hanoi University of Science \\
\normalsize \it 334,Nguyen Trai Str, Thanh Xuan, Ha Noi, Viet Nam.}
\date{March 2002}
\maketitle

\begin{abstract}
Based on the quantum transport equation for the electron-phonon system, the absorption coefficient of sound (acoustic phonons) by absorption of a laser radiation in quantum wires with parabolic potential is calculated for the case of monophoton absorption and the case of multiphoton absorption. Analytical expressions and conditions for the absorption coefficient of sound are obtained. Differences between the two cases of monophoton absorption and of multiphoton absorption are discussed; numerical computations and plots are carried out for a GaAs/GaAsAl quantum wire. The results are compared with bulk semiconductors and quantum wells.
\end{abstract}

\section{Introduction}
It is well known that when a laser radiation is applied to a material, the number of acoustic phonons inside is varied with time. Studied this phenomenon can lead to new knowledge about the electron-phonon interaction mechanism, especially in low dimensional structures. \\
The theory of amplification of sound (acoustic phonons) by absorption of laser radiation in bulk semiconductors and low dimensional structures was studied in a number of papers \cite{bau1,bau2,bau3,ep1,ep2,zhao}. In \cite{bau3,ep1,ep2} the problem was restricted to bulk semiconductor with monophoton absorption and multiphton absorption processes. In \cite{bau1,bau2} the problem was considered for quantum wells with monophoton and multiphoton absorption processes, or in a quantizing magnetic field that do not have the restriction conditions of  \cite{zhao}.\\
In this paper we study theoretically the amplification of sound by absorption of laser radiation in quantum wires with the elliptic cross section using the assymmetric parabolic confining potential:
\[
V(x,y)=\frac{m^\ast}{2}(\Omega_{x}^{2}x^{2}+\Omega_{y}^{2}y^{2}),
\]
Here $\Omega_x,\Omega_y$ are the effective frequencies of the potential, $m^\ast$ is the effective electron mass. The energy of electron in this model has the form \cite{prb61}:
\[
\varepsilon_{n,l}(p_z)=\sqrt{\frac{\Omega_x}{2}}(n+\frac{1}{2})+\sqrt{\frac{\Omega_y}{2}}(l+\frac{1}{2})+\frac{p_z^2}{2m^\ast}.
\]
Effect of scattering on the impurities and the reflection of electron mode from boundary between the wire and the electron reservoir are not considered.

\section{The general formula}
The Hamiltonian for electron-phonon system in a quantum wire in external field can be written as:
\begin{multline} \label{E:pt1}
H(t)=\sum_{n,l,\vec{k}}\varepsilon_{n,l} \left( \vec{k}-\frac{e}{c}\vec{A}(t) \right)+\sum_{\vec{q}} \omega_{\vec{q}}b^{+}_{\vec{q}}b_{\vec{q}}+\\
+\sum_{n,l,n',l',\vec k,\vec q} C_{n,l,n',l'}(\vec q)a^{+}_{n,l,\vec{k}+\vec{q}}a_{n',l',\vec{k}}(b^{+}_{-\vec{q}}+b_{\vec{q}}),
\end{multline}
where $a_{n,l,\vec{k}}^{+}$ and $a_{n,l,\vec{k}}$ ($b_{\vec{q}}^+$ and $b_{\vec{q}}$ ) are the creation and annihilation operators of electron (phonon); $\vec{k}$ is the electron wave vector (along the wire's axis: z axis); $\vec{q}$ is the phonon wave vector; $C_{n,l,n',l'}$is the interaction constant of electron-acoustic phonon scattering; 
$\vec{A}(t)=\frac{c}{\Omega}E_0 cos(\Omega t)$ is the potential vector, depend on the external field.
Process using the method in \cite{bau1,ep1,ep2,bau4}, we obtain the quantum transport equation for acoustic phonons in quantum wires:
\begin{multline}\label{E:pt2}
\frac{\partial}{\partial t}\langle b_{\vec{q}} \rangle_t+i\omega_{\vec{q}} \langle b_{\vec{q}} \rangle_t=-\sum_{n,l,n',l'}|C_{n,l,n',l'}(\vec{q})|^2\sum_{\vec{k}}\left( n_{n,l}(\vec{k}-\vec{q})-n_{n',l'}(\vec{k}) \right) \times\\
\times\int^{t}_{-\infty}\left<b_{\vec{q}} \right>_{t_1}\sum^{\infty}_{\nu,s=-\infty}J_\nu \left( \frac{\lambda}{\Omega}\right) J_s \left( \frac{\lambda}{\Omega}\right)\times\\
\times exp\left( i[\varepsilon_{n',l'}(\vec{k})-\varepsilon_{n,l}(\vec{k}-\vec{q})](t_1-t)-i\nu \Omega t_1+is\Omega t\right) dt_1
\end{multline}
Where the symbol $<x>_t$ means the usual thermodynamics average of operator x; $n_{n,l}(\vec{k})$ is the distribution function of electron; $\lambda = \frac{e\vec{q}\vec{E_0}}{m^\ast \Omega}$. All formulae are written in units where $\hbar=1$\\
Using Fourier transformation, with $\delta\rightarrow +0$, we have the dispersion equation:
\begin{multline}
\omega-\omega_{\vec{q}}=-\pi \sum_{n,l,n',l'}|C_{n,l,n',l'}(\vec{q})|^2 \sum_{\vec{k}}[n_{n',l'}(\vec{k})-n_{n,l}(\vec{k}-\vec{q})]\times\\
\times\sum^{\infty}_{\nu=-\infty}J^2_\nu \left(\frac{\lambda}{\Omega}\right) \frac{1}{\varepsilon_{n',l'}(\vec{k})-\varepsilon_{n,l}(\vec{k}-\vec{q})-\omega_{\vec{q}}-\nu \Omega-i \delta}.\notag
\end{multline}
From this we obtain the general formula for absorption coefficient of acoustic phonons in quantum wires:
\begin{multline} \label{E:pt3}
\alpha(\vec{q})=-\pi \sum_{n,l,n',l'}|C_{n,l,n',l'}(\vec{q})|^2 \sum_{\vec{k}}[n_{n',l'}(\vec{k})-n_{n,l}(\vec{k}-\vec{q})]\times\\
\times \sum^{\infty}_{\nu=-\infty}J^2_\nu \left(\frac{\lambda}{\Omega}\right) \delta \{ \varepsilon_{n',l'}(\vec{k})-\varepsilon_{n,l}(\vec{k}-\vec{q})-\omega_{\vec{q}}-\nu \Omega \},
\end{multline}
where $\delta (x)$ is the Dirac equation.

\section{The amplification of sound}
In the following, we will calculate the absorption coefficient of sound by absorption of laser radiation in quantum wires in the two cases: monophoton absorption process and multiphoton absorption process.

\subsection{Monophoton absorption process}
From Eq.\eqref{E:pt3}, assuming the parameter in Bessel function is small enough: $\lambda\ll\Omega$ and the electron is non-degenerate, we obtain the expression for the sound absorption coefficient:
\begin{multline} \label{E:pt4}
\alpha(\vec{q})=\frac{m^\ast n_0 \lambda^2}{4q\Omega^2}\sum_{n,l,n',l'}|C_{n,l,n',l'}(\vec{q})|^2 \Lambda_{n,l,n',l'}(\vec{q},\Omega)\times\\
\times exp\left( -\frac{\beta}{2}\left( \sqrt{\frac{\Omega_x}{2}}\left(n'+\frac{1}{2}\right)+ \sqrt{\frac{\Omega_y}{2}}\left(l'+\frac{1}{2}\right)\right)-\frac{m^\ast \beta}{2q^2}(a^2+\Omega^{2}) \right),
\end{multline}
where $n_0$ is the electron density,
\[
a=\sqrt{\frac{\Omega_x}{2}}(n-n')+\sqrt{\frac{\Omega_y}{2}}(l-l')+\omega_{\vec{q}}+\frac{q^2}{2m^\ast},
\]
and
\[
\Lambda_{n,l,n',l'}(\vec{q},\Omega)=e^{\frac{\beta \omega_{\vec{q}}}{2}}\left( e^{-\frac{\beta m^\ast a \Omega}{q^2}+\frac{\beta \Omega}{2}}sh\left( \beta\frac{\omega_{\vec{q}}+\Omega}{2} \right)+e^{\frac{\beta m^\ast a \Omega}{q^2}-\frac{\beta \Omega}{2}}sh\left( \beta\frac{\omega_{\vec{q}}-\Omega}{2} \right) \right).
\]
Due to the $\delta$ function in \eqref{E:pt3}, the momentum must satisfy:
\begin{equation} \label{E:pt5}
\vec{k}\geq \mid \frac{\vec{q}}{2}+\frac{m^\ast}{\vec{q}}\left(\sqrt{\frac{\Omega_x}{2}}(n-n')+\sqrt{\frac{\Omega_y}{2}}(l-l')+\omega_{\vec{q}} \right)+\frac{\Omega m^\ast}{\vec{q}}\mid.
\end{equation}
From Eq.\eqref{E:pt4}, when $\omega_{\vec{q}} \ll\Omega$, we see that the absorption coefficent is negative, or it is the amplification of sound: $\alpha(\vec{q})<0$:
\begin{multline}\label{E:pt6}
\alpha(\vec{q})=-\frac{m^\ast n_0 \lambda^2}{2\vec{q}\Omega^2}\sum_{n,l,n',l'}|C_{n,l,n',l'}(\vec{q})|^2 exp\left( \frac{\beta \omega_{\vec{q}}}{2}-\frac{m^\ast \beta \Omega^2}{2q^2}\right)\times\\
\times exp\left[ -\frac{\beta}{2}\left( \sqrt{\frac{\Omega_x}{2}}\left(n'+\frac{1}{2}\right)+ \sqrt{\frac{\Omega_y}{2}}\left(l'+\frac{1}{2}\right)\right)-\frac{m^\ast \beta a^2}{2q^2}\right]\times\\
\times sh\left(\frac{\beta \Omega}{2} \right)sh\left( \frac{\beta m^\ast \Omega}{q^2} \left( \omega_{\vec{q}}+\sqrt{\frac{\Omega_x}{2}}(n-n')+ \sqrt{\frac{\Omega_y}{2}}(l-l')\right)\right).
\end{multline}

\subsection{Multiphoton absorption process}
Assume the parameter in Bessel function is large: $\lambda \gg \Omega$, use the approximate formula \cite{son,sholimal}:
\[
\sum_{\nu}J^2_{\nu} \left( \frac{\lambda}{\Omega} \right)\delta(E-\nu 
\Omega)=\frac{\theta(\lambda^2-E^2)}{\pi \sqrt{\lambda^2-E^2}},
\]
with
\begin{equation}
 \theta =
\begin{cases}
	1,\quad	&\text{if $x>0$}	\\
	0,\quad	&\text{if $x<0$}\notag
\end{cases}.
\end{equation}
From the general formula for absorption coefficient of sound \eqref{E:pt3}, we have:
\begin{multline}\label{E:pt7}
\alpha(\vec{q})=\frac{\pi^{\frac{3}{2}}n_0 m^{\ast}}{2q}e^{-\frac{\beta m^{\ast}\lambda^2}{2q^2}}\sum_{n,l,n',l'}|C_{n,l,n',l'}(\vec{q})|^2\\ \sum^{\infty}_{\nu=0}\frac{\Gamma \left( \nu+\frac{1}{2}\right)}{\nu !} \left( \chi_{n,l}\left( -\frac{q^2}{2m^{\ast}}\right)-\chi_{n',l'}\left(\frac{q^2}{2m^{\ast}}\right) \right),
\end{multline}
where:
\begin{multline*}
\chi_{n,l}(x)=exp\left[\frac{\beta}{2}\left(\sqrt{\frac{\Omega_x}{2}}\left(n'+\frac{1}{2}\right)+\sqrt{\frac{\Omega_y}{2}}\left(l'+\frac{1}{2}\right)\right)\right]\times \\
\times exp\left[-\frac{\beta m^{\ast}}{2q^2}\left(\sqrt{\frac{\Omega_x}{2}}(n-n') +\sqrt{\frac{\Omega_y}{2}}(l-l')+x+\omega_{\vec{q}}\right)^2\right]\times \\
\times\left( \frac{\lambda}{\sqrt{\frac{\Omega_x}{2}}(n-n') +\sqrt{\frac{\Omega_y}{2}}(l-l')+x+\omega_{\vec{q}}} \right)^{\nu}I_{\nu}\left(\frac{\beta m^{\ast}\lambda}{q^2}\right),\notag
\end{multline*}
$I_{\nu}(x)$ is the complex Bessel function of order $\nu$\\.
From the $\theta$ function above, we derive the momentum condition:
\begin{equation}\label{E:pt8}
k<\pm q+\frac{m^{\ast}}{q}\left( \sqrt{\frac{\Omega x}{2}}(n-n')+\sqrt{\frac{\Omega y}{2}}(l-l')+\omega_{\vec{q}} -\mid\lambda \mid\right).
\end{equation}
Note that if $\chi_{n,l}\left(-\frac{q^2}{2m^{\ast}}\right) < \chi_{n',l'}\left(\frac{q^2}{2m^{\ast}}\right) $, then $\alpha(\vec{q})<0$, which mean we have the amplification of sound.

\section{Discussion and numerical results}
Let us begin by discussing the results Eq.\eqref{E:pt4} and Eq.\eqref{E:pt7}, which give the absorption coefficient of sound (acoustic phonons) in quantum wires with parabolic potential in the two cases: monophoton absorption and multiphoton absorption respectively. These expressions are carried out only when the conditions \eqref{E:pt5} and \eqref{E:pt8}, which demonstrate the conservation of energy-momentum, are satisfied. The results are much different from previous work on bulk semiconductors \cite{bau3,ep1,ep2} due to the effect of confinement of electron in low dimensional structures. In quantum wires, the electron's mobility is enhanced, what leads to an unusual behaviour under external stimuli (in this case is the laser radiation). Furthermore, they are also different from results with quantum wells \cite{bau1,bau2}, and quantum wires with cylindrical potential \cite{hung} due to the complexity of the electron energy and wave function. The formula \eqref{E:pt4} demonstrates the relation of the absorption coefficient on laser radiation with order two (the $\lambda = \frac{e\vec{q}\vec{E_0}}{m^\ast \Omega}$ parameter), whereas Eq.\eqref{E:pt7} shows the same relation with order greater than two (since the complex Bessel function $I_{\nu}(x)$ contain $\lambda$). Note that in both cases, in proper conditions, the absorption coefficient is negative, which means the number of phonon increases with time.\\
\begin{figure}
\vspace*{7cm}
\centerline{\hbox to 15cm{\special{eps: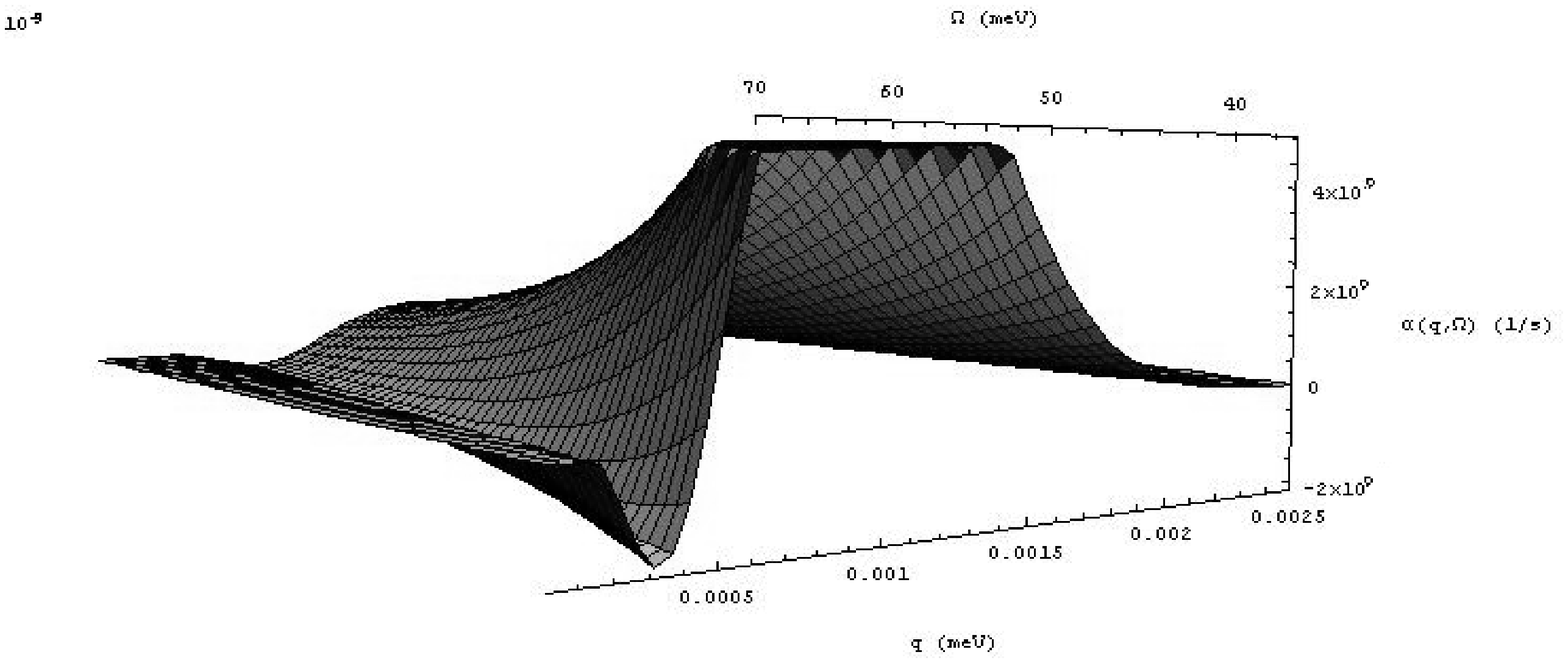 x=15cm y=6cm}}}
\caption{The absorption coefficient of sound versus phonon's wave vector and laser frequency in the case of monophoton absorption; $\beta=k_B T=0.05 meV, \hbar=1$}
\end{figure}
From the obtained results, we plot for the dependence of the absorption of sound $\alpha$ on wave vector $\vec{q}$ and laser frequency $\Omega$ for the case of the monophoton absorption process. The plot is for the case $n \not= n'$, $l \not= l'$, that is for interband absorption, $\beta=k_B T=0.05 meV, r_0=10^{-8}m, m^\ast=0.0067 m_0, m_0$ is the electron mass. The figure shows that the amplification of sound is maximum at $q=0.0005$ (the minimum on the graph).  Compare this plot with those of cylindrical quantum wires, we see that the maximum and the minimum of $\alpha$ happen at the same values of $\vec{q}$, however, the variation of $\alpha$ is slightly difference: it decreases more slowly in the interval of the wavevector from 0.0015 meV to 0.0025 meV.

\section{Conclusion}
In this paper, the amplification of sound (acoustic phonons) by absorption of laser radiation is investigated in the two cases: monophoton absorption process and multiphoton absorption process. The analytical results are difference from previous work, however, the numerical results are similar to those in the case of cylindrical quantum wires. Thus a change in potential does not affect much the electron-phonon scattering in quantum wires.


\begin{thebibliography}{9}
\bibitem{prb61}
 V.A.Geyler, V.A.Margulis,Phys.Rev, B61, 3(2000)1716
\bibitem{hung}
  Nguyen Quoc Hung, Dinh Quoc Vuong, Nguyen Vu Nhan, Nguyen Quang Bau, Pham Thi Nguyet Nga \emph{To be published})
\bibitem{bau1}
  Nguyen Quang Bau, Vu Thanh Tam, Nguyen Vu Nhan, Military science and technology magazine, No 24, 3(1998),38.
\bibitem{bau2}
  Nguyen Quang Bau, Nguyen Vu Nhan, Nguyen Manh Trinh, Proceedings of IWOMS '99, Hanoi 1999, 869.
\bibitem{bau3}
  Nguyen Quang Bau, Nguyen Vu Nhan, Chhoumm Navy, VNU. Journal of Science, Nat.Sci., 15(1999),1.
\bibitem{ep1}
  E.M.Epstein, Radio in Physics, 18(1975),785.
\bibitem{ep2}
  E.M.Epstein, Lett. JEPT, 13(1971)511.
\bibitem{zhao}
  Peiji Zhao, Phys. Rev., B49(1994)13589.
\bibitem{bau4}
  Nguyen Quang Bau, Nguyen Van Huong, Journal of Science of HSU, Physics, 31990)8.
\bibitem{son}
  Nguyen Hong Son, Shmelev G.M., Epstein E.M., Izv. VUZov USSR, Physics, 5(1984)19.
\bibitem{sholimal}
  L.Sholimal, \emph{Tunnel effects in semiconductors and applications}, Moscow, 1974
\end{thebibliography}
\end{document}